# Grain refinement of stainless steel in ultrasound-assisted additive manufacturing

C. J. Todaro[1], M. A. Easton[1], D. Qiu[1], M. Brandt[1], D. H. StJohn[2], M. Qian[1,*]

[1]Centre for Additive Manufacturing, School of Engineering, RMIT University, Melbourne, VIC 3000, Australia.

[2]Centre for Advanced Materials Processing and Manufacturing (AMPAM), School of Mechanical and Mining Engineering, The University of Queensland, St Lucia, QLD 4072, Australia.

[*]Corresponding author: M. Qian, School of Engineering, RMIT University, Melbourne, VIC 3000, Australia.

E-mail address: ma.qian@rmit.edu.au (M. Qian)

**Abstract**

Metals and alloys fabricated by fusion-based additive manufacturing (AM), or 3D printing, undergo complex dynamics of melting and solidification, presenting challenges to the effective control of grain structure. Herein, we report on the use of high-intensity ultrasound that controls the process of solidification during AM of 316L stainless steel. We find that the use of ultrasound favours the columnar-to-equiaxed transition, promoting the formation of fine equiaxed grains with random crystallographic texture. Moreover, the use of ultrasound increases the number density of grains from 305 mm$^{-2}$ to 2748 mm$^{-2}$ despite an associated decrease in cooling rate and temperature gradient in the melt pool during AM. Our assessment of the relationship between grain size and cooling rate indicates that the formation of crystallites during AM is enhanced by ultrasound. Furthermore, the use of ultrasound increases the amount of constitutional supercooling during solidification by lowering the temperature gradient in the bulk of the melt pool, thus creating an environment that favours nucleation, growth, and survival of grains. This new understanding provides opportunities to better exploit ultrasound to control grain structure in AM-fabricated metal products.

**Keywords:** Additive manufacturing, 3D printing, Grain refinement, Ultrasound, Ultrasonic treatment, Steel.

**1. Introduction**

Additive manufacturing (AM), or 3D printing, enables the ability to create complex products that can be mass customised [1-3]. A barrier to widespread implementation of AM of metals is the common occurrence of anisotropic properties in printed products, which is closely associated with coarse columnar grains that grow along the build direction in most commercial alloys [4, 5]. Promoting the columnar-to-equiaxed transition (CET) in AM-fabricated alloys can remove issues with anisotropic properties [6] and optimize combinations of strength, ductility, and toughness [7]. However, the low temperature gradients ($G$) required to form equiaxed grains in many alloys are often difficult to achieve during AM based on established maps of solidification [8-10]. Hence, further advances in AM machines are required to ensure high added-value printed products for the widespread adoption of AM in industrial sectors. In that regard, AM processes are deemed to continue their evolution on all fronts. For example, Todaro et al. [11] have recently demonstrated the use of high-intensity ultrasound to control the grain structures of AM-fabricated Ti-6Al-4V and Inconel 625. The process enabled clear CET leading to fine and equiaxed grains. As a result, the attendant grain structure showed substantially reduced structural anisotropy with noticeably improved tensile strengths [11].

During ultrasound-assisted AM, metal products are developed layer-wise on a vibrating probe by directly feeding metal powder into a melt pool created by a moving laser beam. Meanwhile, high-intensity ultrasound irradiates the melt pool, which remains molten for only about 0.01-0.1 s before solidifying, driving mechanical and physicochemical effects. The primary effect is acoustic cavitation, namely, the formation, growth, and collapse of bubbles in a liquid medium [12], which occurs instantly in molten metallic alloys (~0.00003 s), supported by studies using *in situ* synchrotron X-ray imaging [13]. Acoustic cavitation creates profound energy-matter interactions, with hot spots inside bubbles up to ~5000 °C, pressures up to ~$10^5$ kPa, and heating and cooling rates at ~$10^{10}$ °C $s^{-1}$ [14]. Such effects are essential for refinement of grain structure by ultrasound [15-17], through inducing fragmentation [13, 18] and/or enhancing nucleation of grains [19, 20]. Up till now, the alloys fabricated by ultrasound-assisted AM include Ti-6Al-4V [11], Inconel 625 [11], Al-12Si [21], and Ti-TiB composites [22]. In the last case [22], the use of ultrasound reduced porosity and improved the distribution of reinforcement but resulted in refinement of grain structure at the same time.

In this research, we focus on control of grain structure during AM of metals by extending the ultrasound-assisted AM process to 316L stainless steel, which is widely used in various industrial sectors due to its excellent corrosion resistance, formability, and affordability. AM-fabricated 316L stainless steel is typically composed of columnar grains, leading to anisotropic properties [23, 24]. The addition of sufficient foreign nucleating agents, such as oxides, sulphides, or nitrides [25], can realise CET for improved and consistent properties. However, the nucleants risk facilitating corrosion by pitting in stainless steels [26, 27]. Moreover, such nucleants could agglomerate to form clusters, which can entail degradation in the damage tolerance of products in critical applications. We show that ultrasound-assisted AM can avoid these latent issues and therefore holds promise to produce fine-grained 316L stainless steel for improved and consistent properties without compromising its resistance to corrosion.

## 2. Experiments

Gas-atomised 316L stainless steel powder (45-90 µm) was used to produce cuboid samples with dimensions of 10 mm × 10 mm × 8 mm (length × width × height) by laser directed energy deposition (DED; Trumpf, TruLaser Cell 7020). The sample without using ultrasound was printed on a 4140 steel plate with a laser power of 300 W, laser spot size of 0.61 mm, scan velocity of 10 mm s$^{-1}$, and overlap ratio of 70%. For the sample using ultrasound, an ultrasound processor (Sonic Systems, L500; 20 kHz, 500 W consumed power) together with an ultrasound sonotrode made of 4140 stainless steel (25 mm diameter, 30 µm amplitude of vibration) were used to introduce ultrasound into the melt pool (Fig. 1). An ultrasonic frequency of 20 kHz was selected and considered to be suitable because: (i) the threshold for cavitation starts to increase rapidly with an increasing frequency above 20 kHz [16], (ii) the maximum power output will be compromised at higher frequencies, and; (iii) noise will intensify at lower frequencies (<16 kHz), which is unfriendly. The sample was directly printed on the vibrating sonotrode using the same parameters specified above but with reduced laser power since the ultrasound provides additional input power to the melt in the form of acoustic energy. We first calculated the extra power transmitted to the melt pool by ultrasound, which gives ~125 W (~25% of consumed power [16]). Then we compared with experimental studies and finalised a reduced laser power of 200 W for AM of near-defect-free 316L stainless steel. Samples were printed using linear bi-directional scans with a rotation of 0 and 90° for subsequent layers.

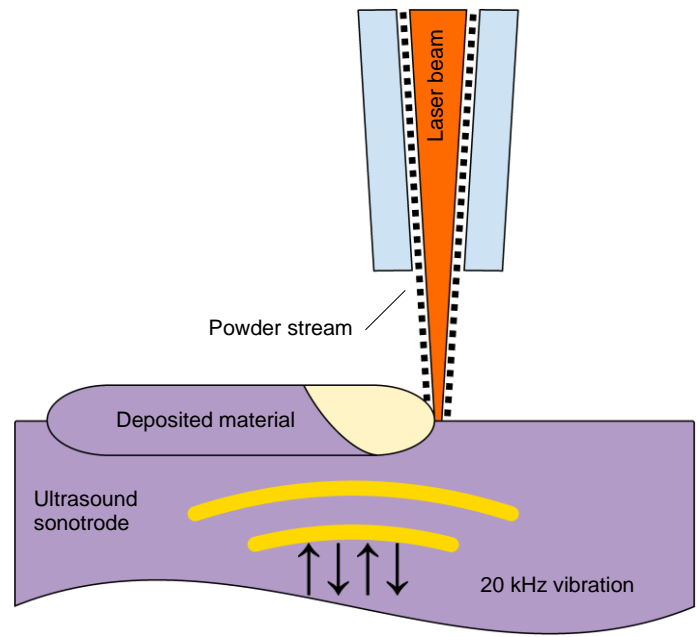

**Fig. 1** Ultrasound-assisted AM. Schematic of AM by laser DED onto an ultrasound sonotrode vibrated at 20 kHz. Adapted from Ref. [11] under the terms of the Creative Commons Attribution 4.0 International (CC BY 4.0) license.

As-printed samples were sectioned along the build direction and prepared for characterisation of microstructure by standard techniques with final polishing by 0.04 µm colloidal silica suspension. Porosity and cracks on the polished sections of the samples were detected by optical microscopy (Leica, DM2500). Grain structure was examined using a scanning electron microscope (JEOL, JSM-7200F) equipped with an electron backscattered diffraction (EBSD) detector (Oxford Instruments, NordlysMax$^2$). The operating parameters used during EBSD were an accelerating voltage of 20 kV, probe current of 16 nA, step size of 1.0 µm for the sample without ultrasound and 0.5 µm for the sample with ultrasound, the working distance of 15 mm, and sample-tilt angle of 70°. The data obtained by EBSD was interpreted using software Channel 5 (Oxford Instruments HKL, Abingdon, UK).

## 3. Results

### 3.1 Formation of defects

To evaluate the effect of ultrasound on the formation of defects in AM-fabricated 316L stainless steel, the samples fabricated without and with the assistance of ultrasound were examined by optical microscopy, as shown in Fig. 2. Both samples are nearly fully dense. The area fraction of porosity on the entire section of each sample approaches 0.01 area% based on measurements using thresholding of the optical microscopy images. Observations made from two other sections per sample are similar. In fact, excluding the sample peripheries, the sample fabricated by ultrasound-assisted AM contains fewer small pores in the bulk of the sample (Fig. 2b), which tends to agree with Ref. [22] that applying ultrasound during AM can reduce porosity. As can be seen from Fig. 2, the side faces of 316L stainless steel are more rugged for the sample with ultrasound, consistent with our previous work on Ti-6Al-4V (see Fig. 2 in Ref. [11]). It is plausible that the use of ultrasound alters the shape of the melt pool, particularly at the periphery of the sample, which deserves further investigation. Nonetheless, these results indicate that near-defect-free 316L stainless steel can be achieved by ultrasound-assisted AM.

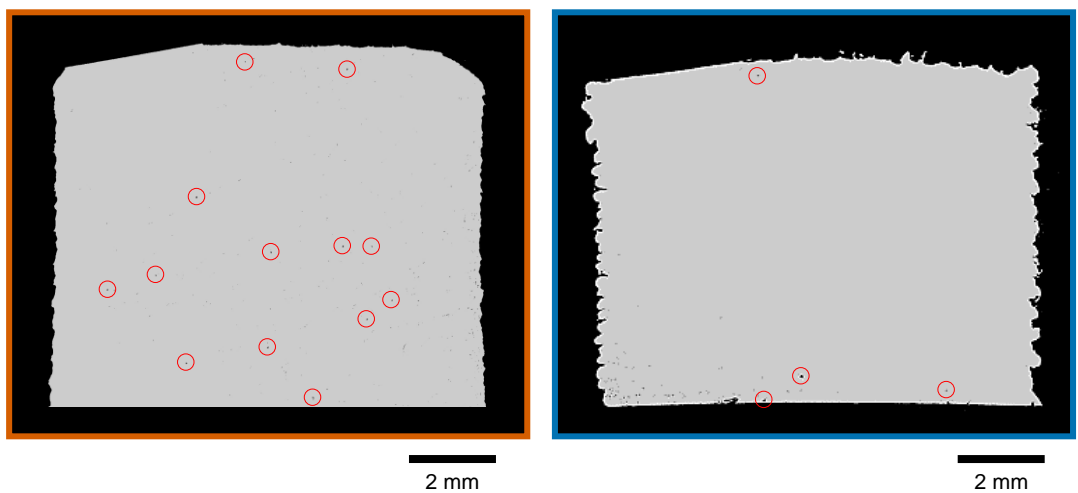

**Fig. 2** Defects in AM-fabricated 316L stainless steel without and with the assistance of ultrasound. **a**, **b** Optical microscopy images of the polished sections of the samples without (**a**) and with (**b**) ultrasound. Major pores are circled.

### 3.2 Formation of grain structure

The distribution of grain orientation and grain size in AM-fabricated 316L stainless steel samples with and without the assistance of ultrasound were characterized by EBSD. Fig. 3a shows the inverse pole figure (IPF) map of the sample without ultrasound, where the grains display irregular morphology, with slight elongation along the build direction. Dotted lines sketch approximate envelopes of the melt pools in layers where the direction of the laser velocity ($\vec{V_L}$) is transverse (that is, the *y*-direction).

Boundaries of the melt pools are determined by assuming that columnar grains usually grow nearly perpendicularly from the bottom of each melt pool [28]. As revealed, many columnar grains of ~50-80 µm in width and ~250 µm in length are almost normal to the fusion boundary at the bottom of each melt pool.

With the assistance of ultrasound, the sample exhibits many fine (~15 µm) nearly equiaxed grains (Fig. 3b). The depth of the melt pool is ~260 µm with ultrasound vs. ~335 µm without ultrasound, indicating that ultrasound has modified the geometry of the melt pool. We assessed the influence of ultrasound on the refinement of grains by examining changes in the number density of grains (number of grains per unit area), which is closely linked to nucleation [29]. The number density of grains based on the high-angle grain boundaries (Fig. 3c, d) is 305 $mm^{-2}$ without ultrasound vs. 2748 $mm^{-2}$ with ultrasound. Such a pronounced increase in number density indicates that ultrasound plays a key role in generating nuclei or crystallites during the solidification of 316L stainless steel when processed by AM.

The CET event may occur during the solidification of a moving melt pool in AM processes, as observed both by experiment [30] and simulation [31]. A zone of columnar grains can exist at the bottom of melt pools, which can transition into a zone of equiaxed grains towards the top surface of the melt pool. We made measurements of the length of the columnar zone along the build direction in seven transverse melt pools per sample using IPF maps obtained by EBSD (see the dashed lines in Fig. 3a, b for examples of determining the CET event). The average length of the columnar zone is reduced from 202 µm (±18 µm standard deviation) without ultrasound to 78 µm (±9 µm standard deviation) with ultrasound, indicating that using ultrasound during AM encourages the CET.

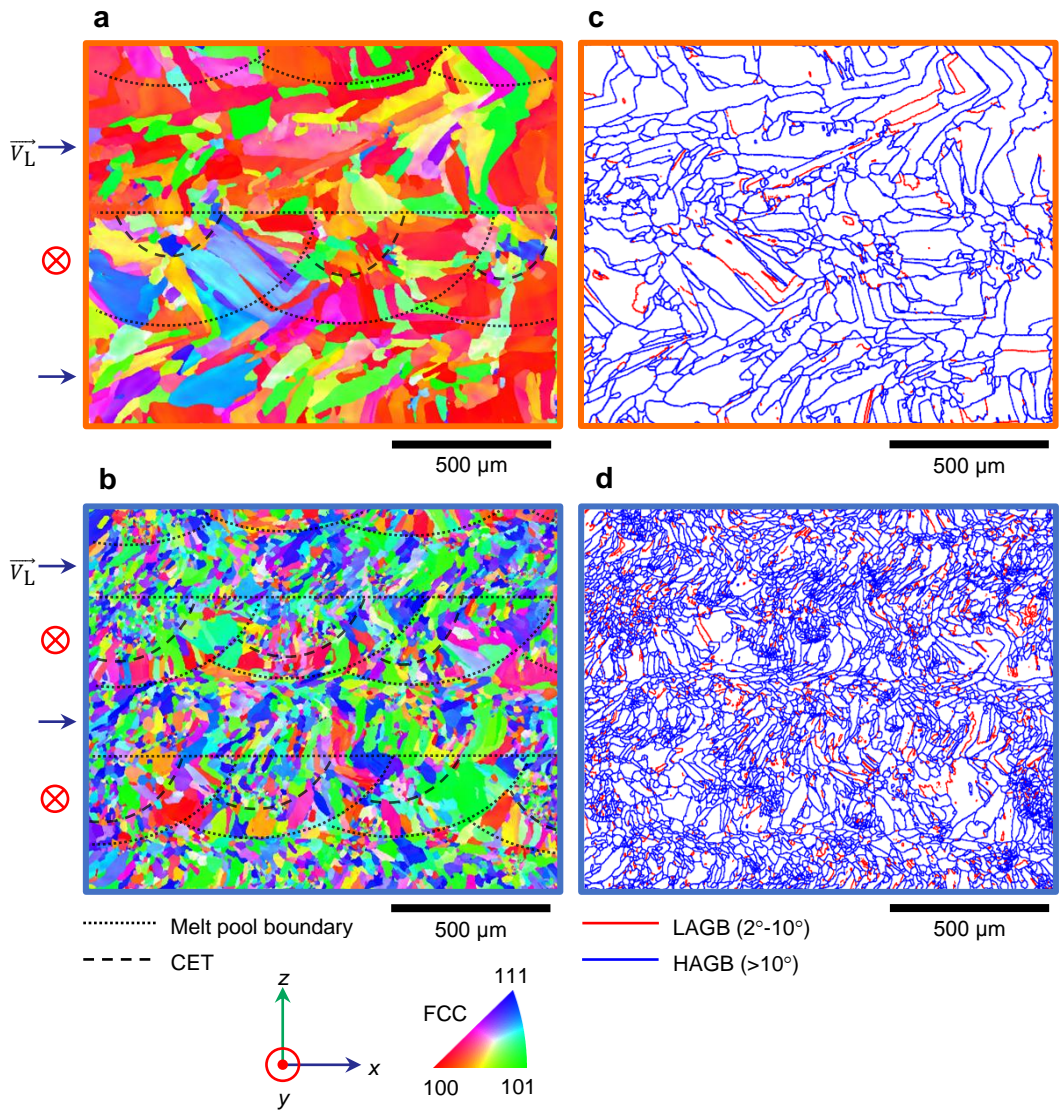

**Fig. 3** Control of grain structure in AM-fabricated 316L stainless steel by ultrasound. **a**, **b** IPF maps along the build direction (*z*) obtained by EBSD showing the orientation of grains in samples without (**a**) and with (**b**) ultrasound. The dotted and dashed lines indicate the approximate boundaries of the melt pools and CET events, respectively, in layers where $\vec{V}_L$ is transverse (*y*-direction). **c**, **d** Grain boundary maps obtained by EBSD showing high angle grain boundaries (HAGBs) and low angle grain boundaries (LAGBs) in samples without (**c**) and with (**d**) ultrasound. HAGBs are coloured blue and LAGBs are coloured red.

**3.3 Homogeneity of grain structure**

To determine the influence of ultrasound on the homogeneity of grain structure, quantitative measurements of grain size ($d$) and grain aspect ratio ($\phi$) were performed from maps of area 1.85 mm ×1.85 mm obtained by EBSD for each sample. The use of ultrasound to AM of 316L stainless steel reduces the grain size $d$ from 52 ± 39 µm to 16 ± 12 µm and the aspect ratio $\phi$ from 2.7 ± 1.6 to 2.0 ± 1.0, as shown in Fig. 4a, b. By defining equiaxed grains with $\phi$ <2.5, near equiaxed grains with 2.5≤ $\phi$ <3.3, and columnar grains with $\phi$ ≥3.3 (after Ref. [32]), by using ultrasound, the frequency of equiaxed grains (defined as $\phi$ <2.5) increases by 21% while the frequency of columnar grains (defined as $\phi$ ≥3.3) decreases by 52%. In that regard, ultrasound-assisted AM culminates in the replacement of many columnar grains with fine equiaxed grains. These observations confirm that the structural homogeneity in AM-fabricated 316L stainless steel improves by the use of ultrasound. Also, The strengthening effect of grain refinement for 316L stainless steel is well recognized [33, 34]. In that regard, the current work offers a good indication of the potential of ultrasound in AM-fabricated 316L stainless steel.

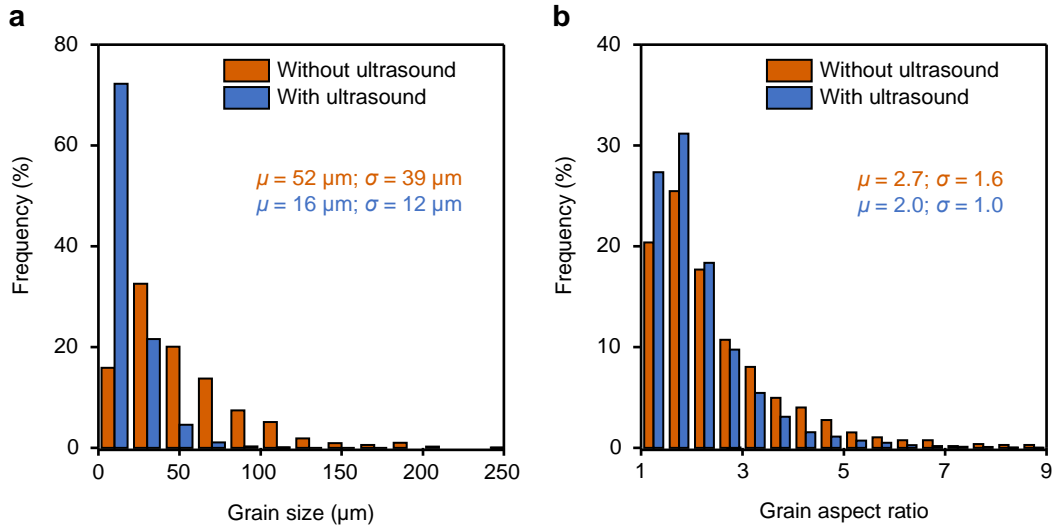

**Fig. 4** Quantitative measurement of grain structure in AM-fabricated 316L stainless steel samples without and with ultrasound. (**a**, **b**) Histograms of grain size (**a**) and grain aspect ratio (**b**) for samples with and without the assistance of ultrasound measured using EBSD data. $\mu$, mean; $\sigma$, standard deviation.

For ultrasound-assisted AM, the average grain size is ~16 µm for 316L stainless steel, ~9 µm for Inconel 625 [11], and ~117 µm for Ti-6Al-4V [11]. Alloy constitution is critically important to obtain fine equiaxed grains under a variety of solidification conditions, including ultrasonic [17, 35, 36] and AM [37-39] conditions. In particular, the generation of enough constitutional supercooling (CS) ahead of the solid-liquid interface during solidification is essential to trigger effective refinement of grain structure in metallic alloys. The amount of CS ($\Delta T_{CS}$) is proportional to the growth restriction factor ($Q$) [40], which is related to the degree of supersaturation ($\Omega$), i.e., $\Delta T_{CS} = Q\Omega$ [41]. We have established the value of $Q$ for 316L stainless steel, Inconel 625, and Ti-6Al-4V using the calculated phase diagram (CALPHAD) method [40]. $Q$ is ~134 ºC for stainless steel, ~184 ºC for Inconel 625, and ~6 ºC for Ti-6Al-4V. The resulting values provide insight into the mechanism by which 316L stainless steel and Inconel 625 have about an order of magnitude finer grain sizes than Ti-6Al-4V when fabricated by ultrasound-assisted AM.

**3.4 Crystallographic texture**

To assess the changes in crystallographic texture by ultrasound, we constructed pole figures for the {100}, {101}, and {111} crystal plane families using the data obtained by EBSD, and the results are shown in Fig. 5. The poles are given in multiples of uniform distribution (MUD), with a maximum value of MUD equal to 1.0 representing a random texture. Without ultrasound, the AM-fabricated 316L stainless steel exhibits a clear crystallographic texture with a maximum value of MUD equal to 3.9 (Fig. 5a). More specifically, each pole figure shows typical patterns of the preferred cube texture component {001}<100> (Fig. 5a), indicating that the <100> crystallographic directions of many of the grains are aligned with the reference directions of the sample (*x*, *y,* and *z*). With ultrasound, the maximum value of MUD reduces from 3.9 to 1.7 and the cube texture component {001}<100> is avoided while no other texture component is found (Fig. 5b). The maximum MUD value of 1.7 obtained by the use of ultrasound is among the lowest values reported to date for as-printed AM-fabricated 316L stainless steel [28, 42, 43]. These results confirm that the use of ultrasound mitigates preferred texture in AM-fabricated 316L stainless steel by producing grains in stochastic orientations, consistent with texture analyses of Ti- and Ni-based alloys fabricated by ultrasound-assisted AM [11].

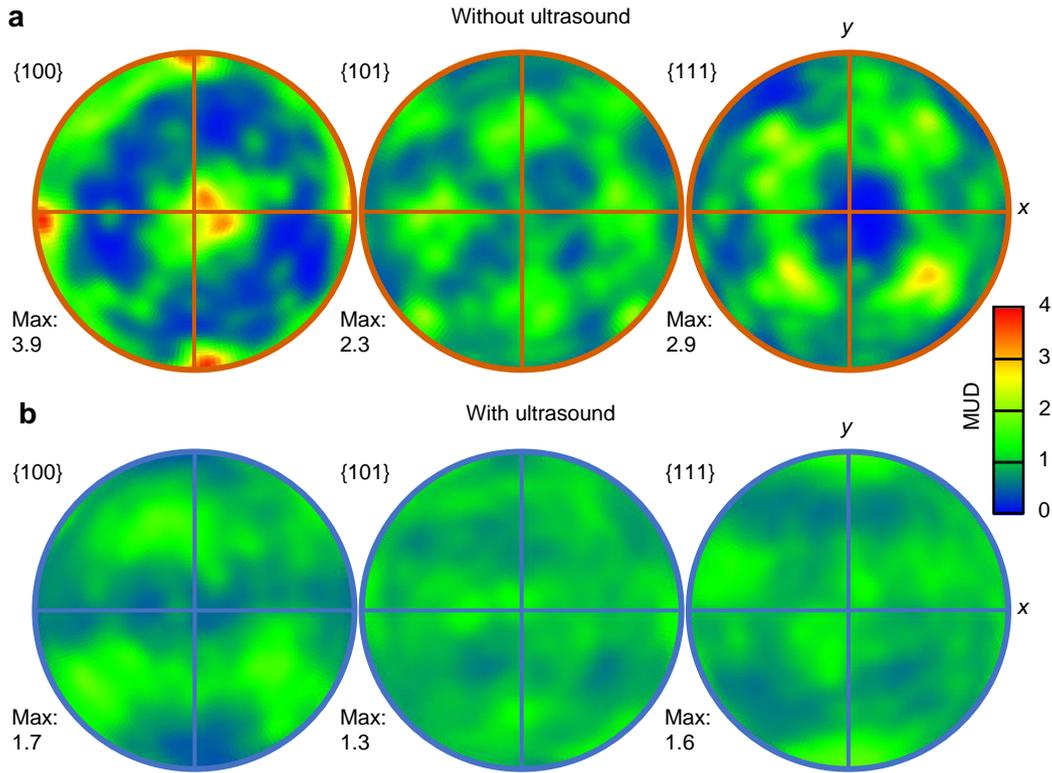

**Fig. 5** Modification of texture in AM-fabricated 316L stainless steel by ultrasound. (**a, b**) The {100}, {101} and {111} contoured pole figures (in MUD) for the samples without (**a**) and with (**b**) ultrasound corresponding to the EBSD maps in Fig. 3. A preferred cube texture component {001}<100> develops without ultrasound, avoided with ultrasound.

## 4. Discussion

### 4.1 Conditions of solidification

To estimate the local conditions of solidification with and without the assistance of ultrasound, we analysed the solidification microstructure of AM-fabricated 316L stainless steel samples. Representative cellular structures of AM-fabricated 316L stainless steel with and without ultrasound are shown in Fig. 6a, b (perpendicular to the build direction). We made 125 measurements of the primary spacing of cells ($\lambda_c$) per sample by the line intercept method from scanning electron microscopy (SEM) backscattered electron images. Fig. 6c compares the primary spacings of cells in samples with and without ultrasound. The average value of primary spacing is 2.0 µm without ultrasound vs. 2.7 µm with ultrasound. The primary spacing of cells $\lambda_c$ can be qualitatively described as a power-law dependence on the cooling rate ($\dot{T}$) according to the relationship [41]:

$$\lambda_c = K|\dot{T}|^n \tag{1}$$

In Eq. 1, $K$ and $n$ are material constants, where $n$ can be assumed to be unitless while the unit of $K$ should accordingly be defined to render the unit of µm to $\lambda_c$. For austenitic stainless steels, $n$ was determined to be -0.33 while $K$ was defined as 80 µm·(ºC s$^{-1}$)$^{0.33}$ according to experimental measurements [44]. Using Eq. 1 and the measurements of average primary spacing, the average value of the cooling rate $\dot{T}$ is estimated to be 7.2 × 10$^4$ ºC s$^{-1}$ without ultrasound and 2.9 × 10$^4$ ºC s$^{-1}$ with ultrasound. The average values of the cooling rate reported here are of the same order of magnitude as those obtained by *in situ* measurements by pyrometry during laser DED of stainless steel [45].

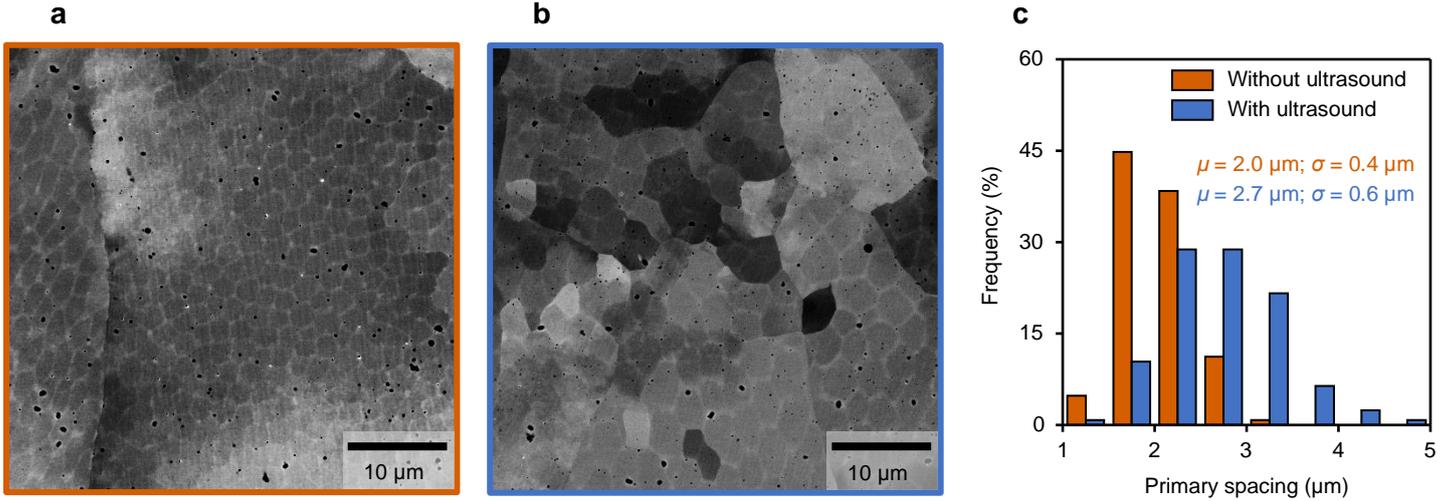

**Fig. 6** Representative cellular structures of AM-fabricated 316L stainless steel with and without the assistance of ultrasound (perpendicular to the build direction). (**a**, **b**) SEM backscattered electron images showing the typical cellular microstructure of samples without (**a**) and with (**b**) ultrasound. (**c**) Histograms of cell spacing for samples with and without ultrasound. $\mu$, mean; $\sigma$, standard deviation.

The cooling rate $\dot{T}$ at a solidifying interface under conditions of unidirectional heat flow is given by [41]:

$$\dot{T} = GV \qquad (2)$$

where $G$ is the temperature gradient and $V$ is the growth rate. As schematically shown in Fig. 7, in a longitudinal section through the centreline of a laser track (that is, the $x$-direction), the growth rate $V$ can be related to the laser velocity ($V_L$) by the relationship [46]:

$$V = V_L \cos\theta \qquad (3)$$

where $\theta$ is the angle between $V$ and $V_L$. Since the grain structure orients itself nearly parallel to the heat flux at high temperature gradients, the angle $\theta$ can be determined experimentally by measuring the amount of rotation between the orientation of grains and the known direction of laser scanning [46]. Consequently, by assuming that the longitudinal laser traces in Fig. 7b, c (extracted from Fig. 3) are sectioned along their centrelines, $\theta$ is ~29° without ultrasound and ~48° with ultrasound. We note that the grains are not perfectly aligned, possibly implying that the melt pools did not reach a steady-state. The laser velocity $V_L$ is 10 mm s$^{-1}$ for both samples. Eq. 3 then gives $V$ = 8.8 mm s$^{-1}$ without ultrasound and $V$ = 6.8 mm s$^{-1}$ with ultrasound. By Eq. 2, the temperature gradient $G$ is ~8.2 × 10$^3$ °C mm$^{-1}$ without ultrasound and ~4.3 × 10$^3$ °C mm$^{-1}$ with ultrasound. The conditions of solidification in AM-fabricated 316L stainless steel with and without the assistance of ultrasound are summarised in Table 1. These results indicate that the use of ultrasound reduces the temperature gradient ahead of the solid-liquid interface during solidification in AM by about 50%.

**Table 1.** The conditions of solidification in AM-fabricated 316L with and without the assistance of ultrasound.

| Condition | Cooling rate, $\dot{T}$ (°C s$^{-1}$) | Growth rate, $V$ (mm s$^{-1}$) | Temperature gradient, $G$ (°C mm$^{-1}$) |
|---|---|---|---|
| Without ultrasound | 7.2 × 10$^4$ | 8.8 | 8.2 × 10$^3$ |
| With ultrasound | 2.9 × 10$^4$ | 6.8 | 4.3 × 10$^3$ |

In addition to post-mortem analysis of microstructure which suggests that ultrasound decreases the temperature gradient, recent modelling of the flow of fluid has shown that acoustic streaming generated by ultrasound established a markedly lowered temperature gradient in a cast alloy Al-2Cu [47]. In this work, given the 0.61 mm laser spot size and the 10 mm s$^{-1}$ scan velocity, the dwell time of the laser beam is 61 ms, which is much longer than the period of ultrasound (0.5 ms), allowing sufficient interactions between the ultrasound and the melt pool. Meanwhile, acoustic streaming is expected to develop immediately during ultrasound-assisted AM due to the small size of the melt pool (~260 μm deep). In this regard, it is reasonable that the use of ultrasound could enhance convection to reduce the temperature gradient in AM. Besides, the heat generated by ultrasonic energy may also decrease the cooling rate of the melt pool during solidification, as demonstrated by the thermal analysis of bulk melts treated with ultrasound [48].

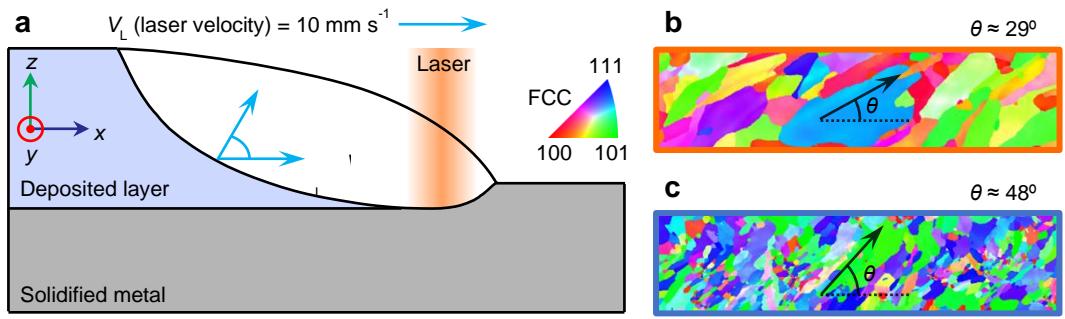

**Fig. 7** The typical form of a melt pool in AM by laser DED. **a** schematic of a longitudinal section through the centerline of a solidifying laser track. **b**, **c** IPF maps along the build direction (*z*) of a longitudinal cut across a laser track without (**b**) and with (**c**) ultrasound.

Experimental data from the literature [49-57] and this work on grain size *d* and primary spacing of cells *λ*$_c$ of AM-fabricated 316L stainless steel are compiled in Table 2. We used this data to plot how grain size *d* varies with cooling rate $\dot{T}$ (calculated using Eq.

1), as shown in Fig. 8. A strong correlation is observed between the grain size $d$ of samples without ultrasound (orange symbols) and the inverse square root of cooling rate $\dot{T}$ of following the equation:

$$d = 10.4 + \frac{1.2 \times 10^3}{\sqrt{\dot{T}}}, \qquad R^2 = 0.91 \qquad (4)$$

Similarly, grain size $d$ was previously demonstrated to be linearly related to the inverse square root of cooling rate $\dot{T}$ in cast Al- [58] and Mg-based [59] alloys, suggesting that the linear relationship may apply to a range of alloys and conditions of solidification.

A striking feature of the plot in Fig. 8 is that AM-fabricated 316L stainless steel with ultrasound (blue symbol) does not follow the grain size-cooling rate relationship revealed for without ultrasound (orange symbols). More specifically, the experimentally measured grain size with ultrasound is about five times smaller than that predicted by Eq. 4 (15 µm vs. 83 µm, respectively). In that regard, the use of ultrasound provides favourable conditions for generating nuclei or crystallites during solidification, thus decreasing the grain size compared without using ultrasound.

**Table 2** Comparison of the grain size and cell spacing of as-built AM-fabricated 316L stainless steel. The grain size is given as the grain width.

| AM process | Grain size (μm) | Cell spacing (μm) | Ref. |
|---|---|---|---|
| Laser powder bed fusion | ~16 | ~0.5 | [49] |
| | ~18 | ~0.6 | |
| | ~19 | ~0.7 | |
| | ~21 | ~0.9 | |
| | ~22 | ~0.6 | |
| | ~25 | ~0.8 | |
| | ~24 | ~1.0 | |
| | ~27 | ~1.3 | |
| Electron beam powder bed fusion | ~76 | ~2.6 | [50] |
| Laser powder bed fusion | ~12 | ~0.4 | [51] |
| Laser powder bed fusion | ~25 | ~1.25 | [52] |
| Laser powder bed fusion | ~15 | ~0.6 | [53] |
| Laser directed energy deposition | ~45 | ~1.8 | [54] |
| Laser powder bed fusion | ~30 | ~0.67 | [55] |
| Laser powder bed fusion | ~15 | ~1.0 | [56] |
| Laser powder bed fusion | ~21 | ~0.4 | [57] |
| Laser directed energy deposition | ~60 | ~2 | This work |
| Laser directed energy deposition (with ultrasound) | ~15 | ~2.7 | This work |

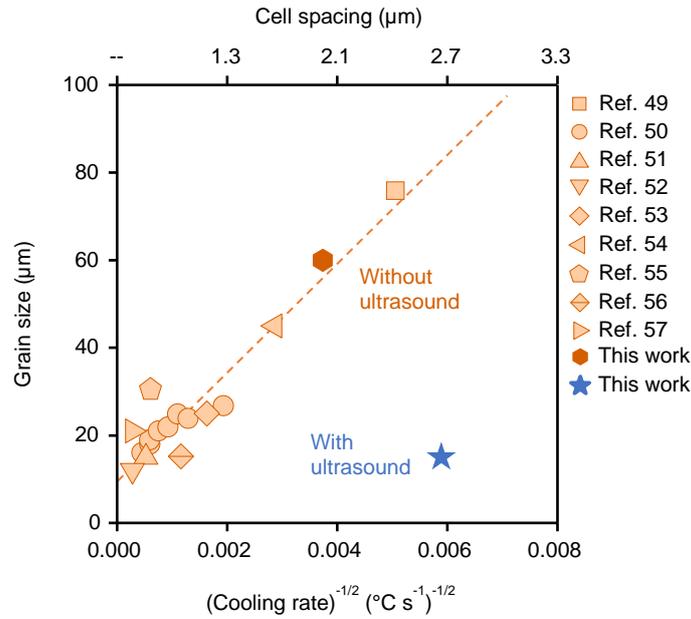

**Fig. 8** The dependence of grain size $d$ on cooling rate $\dot{T}$ in AM-fabricated 316L stainless steel. Grain size $d$ with the inverse square root of cooling rate $\dot{T}$ from the literature (open symbols) [49-57] and this work (solid symbols). The dashed line represents the line of best fit for the samples without ultrasound (orange symbols). The samples without ultrasound reveal a strong linear relationship ($R^2 = 0.91$) while the sample with ultrasound (blue symbol) deviates from this relationship.

### 4.2 Effect of ultrasound on solidification

CS largely controls the development of grain structure in metallic alloys solidified under a variety of conditions [60, 61], including under ultrasonic [17, 35, 36] and AM conditions [37-39]. In particular, the CET requires some growth of columnar grains to generate enough CS to trigger nucleation of equiaxed grains on nucleants of potency $\Delta T_n$, i.e., requiring $\Delta T_{CS} \geq \Delta T_n$. This hypothesis has been verified by carefully designed experimental studies in casting [62] and AM [37] technologies. In this work, the amount of columnar growth required for the CET reduced from 202 µm to 78 µm by the use of ultrasound (Fig. 3), suggesting that ultrasound influences CS.

We schematically illustrate (Fig. 9) the development of the CS zone ahead of a growing grain during the printing of a layer with and without the assistance of ultrasound. Without ultrasound (Fig. 9a), the first event occurs by epitaxial growth on the partially remelted previous layer at time $t_1$. The steep temperature gradient $G$ estimated to be ~$8.2 \times 10^3$ °C mm$^{-1}$ limits $\Delta T_{CS}$ ahead of the growing grain (the yellow region, which is the difference between the equilibrium liquidus temperature, $T_E$, and the actual temperature in the melt, $T_A$), preventing activation of a potent nucleant present ahead of the solid-liquid interface and promoting columnar growth until time $t_3$. At time $t_3$, the columnar growth generates enough $\Delta T_{CS}$ to trigger activation of a potent nucleant, i.e., $\Delta T_{CS} \geq \Delta T_n$, driving nucleation and the CET event.

Solidification under high-intensity ultrasound is different (Fig. 9b). With ultrasound, the first event still occurs by epitaxial growth at time $t_1$. However, the reduced temperature gradient from ~$8.2 \times 10^3$ °C mm$^{-1}$ to ~$4.3 \times 10^3$ °C mm$^{-1}$ by ultrasound increases $\Delta T_{CS}$ (the yellow region). Consequently, the amount of columnar growth required to trigger the activation of a potent nucleant and the CET when $\Delta T_{CS} \geq \Delta T_n$ is significantly reduced (from 202 µm to 78 µm, Fig. 3), occurring earlier at time $t_2$. Also, high-intensity ultrasound plays a key role to produce many initial crystallites near the solid-liquid interface, through cavitation-induced fragmentation [13, 18] and/or cavitation-enhanced nucleation [19, 20] processes. This is corroborated by the observation that low-frequency mechanical vibrations (≤5000 Hz), where cavitation effects are absent, are only able to induce limited refinement of grain structure during laser DED [63]. The larger CS zone facilitated by ultrasound protects the cavitation-generated crystallites from readily re-melting as they move away from the solid-liquid interface. Thus, despite reducing the cooling rate (Table 1), ultrasound creates a solidification environment that favours nucleation, growth, and survival of grains, thus facilitating refinement of grain structure and the CET. This concept can be used to explain the formation of a fine mostly equiaxed structure in AM-fabricated 316L stainless steel with the assistance of ultrasound.

We note that the discussion above assumes that the use of ultrasound has no or negligible effect on the segregation of solute and the equilibrium liquidus temperature $T_E$. In experiments dedicated to testing the development of CS at the solid-liquid interface under a variety of ultrasound conditions, it was unexpectedly found that ultrasound showed little influence on the redistribution of solute up to a level of ultrasound intensity equal to 1700 W cm$^{-2}$ [29]. In that regard, ultrasound may only influence the solute boundary layer to a limited extent over a short timescale and does not appreciably affect the short-range diffusion field of a developing CS zone, which should be verified in future studies.

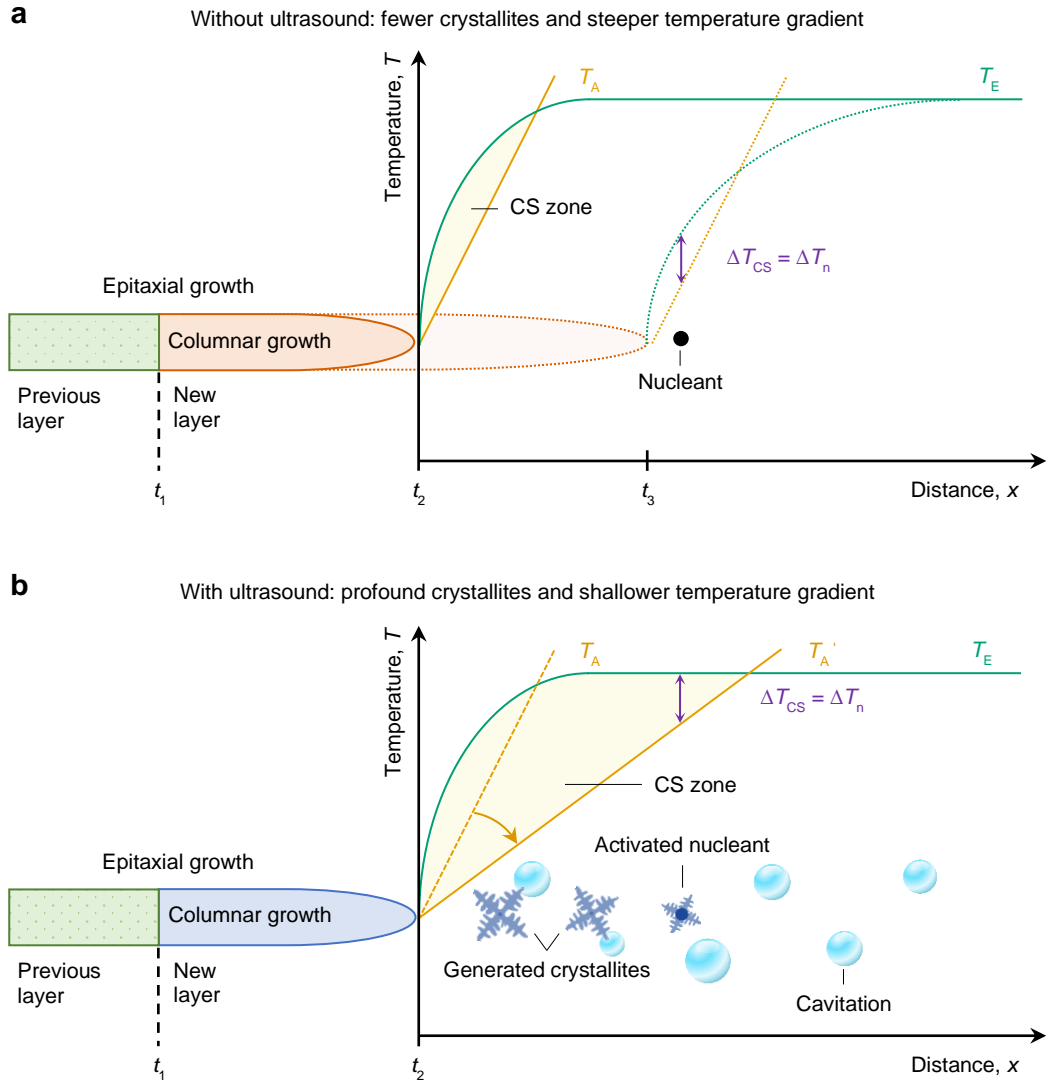

**Fig. 9** Schematic illustration of the CS zone ahead of a growing grain during AM with and without the assistance of ultrasound. **a** Without ultrasound, the formation of grain structure commences by epitaxial columnar growth from the prior layer at time $t_1$. At time $t_2$, the developing $\Delta T_{CS}$ (the difference between the equilibrium liquidus temperature, $T_E$, and the actual temperature in the melt, $T_A$) is less than the undercooling required for nucleation $\Delta T_n$ on a nucleant particle, then columnar growth continues. Columnar growth continues to time $t_3$, where $\Delta T_{CS} = \Delta T_n$, then equiaxed nucleation occurs and the CET is triggered. **b** With ultrasound, the CS zone is larger and longer due to the lowered temperature gradient from $\frac{d\Delta T_A}{dx}$ to $\frac{d\Delta T_A'}{dx}$, triggering nucleation of more grains at time $t_2$. Furthermore, acoustic cavitation generates many initial crystallites that can survive and grow in the CS zone. Hence, ultrasound facilitates

nucleation, growth, and survival of grains, creating conditions for the formation of a fine equiaxed grain structure.

Fine equiaxed grain structures can be obtained in AM by employing alloys with solute elements that generate high values of $Q$ to provide CS, as demonstrated in alloys Ti-W [64] and Ti-Cu [38]. Also, for AM-fabricated alloys with low values of $Q$, e.g., Ti-6Al-4V with $Q \approx 0$ °C, the use of ultrasound can produce fine equiaxed grain structures [11]. 316L stainless steel has a moderate value of $Q$ equal to ~134 °C. Meanwhile, thermal undercooling ($\Delta T_{therm}$), which is always present in AM due to the high cooling rates, provides additional undercooling [65] (curvature undercooling can be neglected since it is generally insignificant across a variety of conditions of solidification [66]). In this work, the total undercooling ($\Delta T_{total} = \Delta T_{CS} + \Delta T_{therm}$) is insufficient to promote significant nucleation for early CET in AM-fabricated 316L stainless steel without the assistance of ultrasound with respect to the nucleating particles that naturally exist in the alloy melt. In contrast, the use of ultrasound generates new crystallites by enabling fragmentation [13, 18] or enhancing nucleation [19, 20] and fundamentally increases the region of $\Delta T_{CS}$ by reducing the temperature gradient (Fig. 9). These combined effects promote refinement of grain structure and earlier CET in the sample fabricated by ultrasound-assisted AM.

## 5. Conclusions

In summary, 316L stainless steel was fabricated by additive manufacturing (AM) with and without the assistance of ultrasound. Without ultrasound, the grain structure mostly consists of 250 μm long columnar grains with a preferred cube texture component

{001}<100>. In contrast, with ultrasound, the columnar-to-equiaxed (CET) event occurs much earlier and the resultant grain structure shows predominantly fine (~15 µm) near equiaxed grains with no preferred texture. Despite a decrease in cooling rate and temperature gradient in the melt pool, the number density of grains increases substantially from 305 mm$^{-2}$ to 2748 mm$^{-2}$ by the use of ultrasound. Also, the grain size of AM-fabricated 316L stainless steel with ultrasound deviates considerably from the grain size-cooling rate relationship revealed for conventional AM-fabricated 316L stainless steel. The increase in the number density of grains can be attributed to the use of ultrasound generating many initial crystallites and facilitating the formation of a larger constitutional supercooling zone due to the lowered temperature gradient. Both these factors promote the CET and the formation of a fine equiaxed grain structure.

**Acknowledgements**

This research work was supported by the Australian Research Council (ARC) Discovery Projects DP150104719, DP160100560 and DP140100702 and the ExoMet Project co-funded by the European Commission's 7th Framework Programme (contract

FP7-NMP3-LA-2012-280421), by the European Space Agency and by the individual partner organizations. We thank both the Microscopy and Microanalysis Facility (RMMF) and the Advanced Manufacturing Precinct (AMP) at RMIT University for their facilities and technical assistance.